\documentclass[%
 reprint,twocolumn, notitlepage,
 amsmath,amssymb,
 aps,
]{revtex4-2}

\usepackage{lipsum}

\usepackage{graphicx}
\usepackage{dcolumn}
\usepackage{bm}
\usepackage{hyperref}

\usepackage{tikz}
\hypersetup{%
        colorlinks=true,
        linkcolor=blue,
        citecolor=blue,
        urlcolor=blue,
      }

\newcommand{\ket}[1]{|#1\rangle}
\newcommand{\bra}[1]{\langle#1|}

\begin{document}


\title{Parametric-resonance entangling gates with a tunable coupler}
\author{Eyob A. Sete}
\thanks{Corresponding author, eyob@rigetti.com}
\affiliation{%
Rigetti Computing, 775 Heinz Avenue, Berkeley, California 94710, USA
}%
\author{Nicolas Didier}
\affiliation{%
Rigetti Computing, 775 Heinz Avenue, Berkeley, California 94710, USA
}%
\author{Angela Q. Chen}
\affiliation{%
Rigetti Computing, 775 Heinz Avenue, Berkeley, California 94710, USA
}
\author{Shobhan Kulshreshtha}
\affiliation{%
Rigetti Computing, 775 Heinz Avenue, Berkeley, California 94710, USA
}%
\author{Riccardo Manenti}
\affiliation{%
Rigetti Computing, 775 Heinz Avenue, Berkeley, California 94710, USA
}%
\author{Stefano Poletto}
\affiliation{%
Rigetti Computing, 775 Heinz Avenue, Berkeley, California 94710, USA
}%

\date{\today}
\begin{abstract}

High-fidelity parametric gates have been demonstrated with superconducting qubits via rf flux modulation of the qubit frequency. The modulation however leads to renormalization of the bare qubit-qubit coupling, thereby reducing the gate speed. Here, we realize a parametric-resonance gate, which is activated by bringing the average frequency of the modulated qubit in resonance with a static-frequency qubit while approximately retaining the bare qubit-qubit coupling. The activation of parametric-resonance gates does not depend on the frequency of modulation, allowing us to choose the modulation frequencies and avoid frequency collisions. Moreover, we show that this approach is compatible with tunable coupler architectures, which reduce always-on residual couplings. Using these techniques, we demonstrate iSWAP and CZ gates between two qubits coupled via a tunable coupler with average process fidelities as high as $99.3\%$ and $97.9\%$, respectively. The flexibility in activating parametric-resonance gates combined with a tunable coupler architecture provides a pathway for building large-scale quantum computers.

\end{abstract}

\pacs{Valid PACS appear here}
\maketitle


\section{Introduction}
One of the main challenges of building practical quantum computing architectures is the implementation of high-fidelity entangling gates at scale. When increasing the quantum processor size, issues such as parasitic couplings from neighboring qubits and frequency crowding can limit two-qubit gate performance. Architectures with tunable couplers have been proposed as a way to tune the qubit-qubit coupling to zero at the idle point and thereby eliminate parasitic couplings~\cite{Chen2014,McKay16,Oliver18}, making it possible to achieve fast, high-fidelity two-qubit gates approaching $99.9\%$~\cite{Oliver20,Stehlik21}. This motivated further exploration of different tunable coupler ~\cite{Stehlik21,Sete21} and two-qubit gate ~\cite{Oliver18,Martinis19,Houck19,Oliver20,Dapeng20,Wallraff20,Foxen20} implementations. Other methods can also be used to eliminate unwanted $ZZ$ couplings, for example, quantum optimal control \cite{Goerz2017}. To address the inevitable issue of frequency crowding in large-scale processors, parametric gates have been introduced as a flexible way to activate gates ~\cite{Reagor18,Caldwell18,Didier18,Hong20}. Parametric two-qubit gates are activated by radio frequency modulating one of the qubits to bring the sidebands in resonance with the neighboring qubit. As a result of the modulation, the coupling between the qubits is renormalized by the weights of the sidebands, reducing the effective qubit-qubit coupling and making the gate slower.

In this work, we implement entangling gates via on-resonance parametric modulation between transmon qubits coupled through a tunable coupler. The gates, which we refer to as \emph{parametric-resonance gates}, are activated by bringing the average frequency of the modulated qubit in resonance with the unmodulated qubit instead of using sideband transitions~\cite{McKay16,Reagor18,Caldwell18,Didier18,Hong20,Didier21}. This enables us to approximately retain the bare coupling strength and enact fast entangling gates. Moreover, since the resonance conditions for parametric-resonance gates do not depend on the frequency of modulation, they can be chosen to avoid frequency collisions with sideband resonances. Furthermore, the use of a tunable coupler allows us to minimize the always-on ZZ at the idle bias at which the qubits are parked. The tunable coupler is particularly advantageous for parametric-resonance gates as it makes it possible to work with weakly detuned qubits. Because of this, the gates can be activated with a small modulation amplitude that reduces the dephasing time due to flux noise and minimizes the sensitivity to two-level system defects.

We experimentally demonstrate parametric-resonance iSWAP and controlled-Z (CZ) gates with average process fidelities as high as $99.3\%$ (duration 44 ns) and $97.9\%$ (duration 124 ns), respectively. The iSWAP fidelity is largely limited by qubit coherence, whereas the CZ fidelity is limited by phase error due to the repulsion from higher energy levels of the qubits and coupler. Both gates are implemented by modulating the flux pulse at frequencies higher than the activation frequency of sideband gates to steer away from collisions. The ability to perform native iSWAP and CZ gates can potentially enable efficient compilation of quantum circuits such as MaxCut quantum approximate optimization algorithm on quantum processors with limited connectivity ~\cite{Abrams20,Babbush17,Babbush18,Kivlichan18,Crooks18,OGorman19}.

\section{Parametric-resonance gates} 
Thus far parametric gates are enabled by bringing the sideband of a parametrically modulated qubit in resonance with the frequency of an unmodulated qubit ~\cite{Didier18,Caldwell18,Reagor18,Hong20}. Following these studies, we first present the general theory for the parametric-resonance gate with fixed coupling strengths. We then consider the implementation of the parametric-resonance gate in a tunable coupler architecture . Indexing the states of coupled qubits $|\mathrm{qubit\ 1}, \mathrm{qubit\ 2}\rangle$, the interaction of a modulated frequency-tunable qubit (qubit 2) and an unmodulated qubit (qubit 1) can be described in the interaction picture by the Hamiltonian~\cite{Didier18} 
\begin{align}\label{hamiltonian}
    H_{\rm int} = \sum_{n=-\infty}^{\infty} &g_n [e^{i(2n\omega_p -\Delta) t}|10\rangle
\langle 01| \notag\\
&+\sqrt{2}e^{i(2n\omega_p-\Delta +\eta_2)t}|11\rangle\langle 02| \notag\\
&+ \sqrt{2}e^{i(2n\omega_p-\Delta -\eta_1)t}|20\rangle\langle 11| +\mathrm{H.c.}],
\end{align}
where we have assumed a small modulation amplitude and ignored the flux dependence of the couplings and anharmonicities. Here, $g_n = gJ_n\left(\tilde \omega_2/2\omega_p\right) e^{i2n\phi_p}$ is the renormalized coupling due to the modulation, $g$ is the bare coupling rate, $\tilde \omega_2$ is the qubit frequency excursion during modulation, $\omega_p$ and $\phi_p$ are the modulation frequency and phase of the flux pulse, and $J_n$ is the $n$th-order Bessel function of the first kind. The parameter $\Delta = \overline \omega_2-\omega_1$ represents the detuning between the average frequency $\overline\omega_{2}$ of qubit 2 and the frequency of the first qubit, $\omega_1$. Lastly, $\eta_k$ is the absolute value of the anharmonicity of the qubits. 
\begin{figure}
    \centering
    \includegraphics[width=\linewidth]{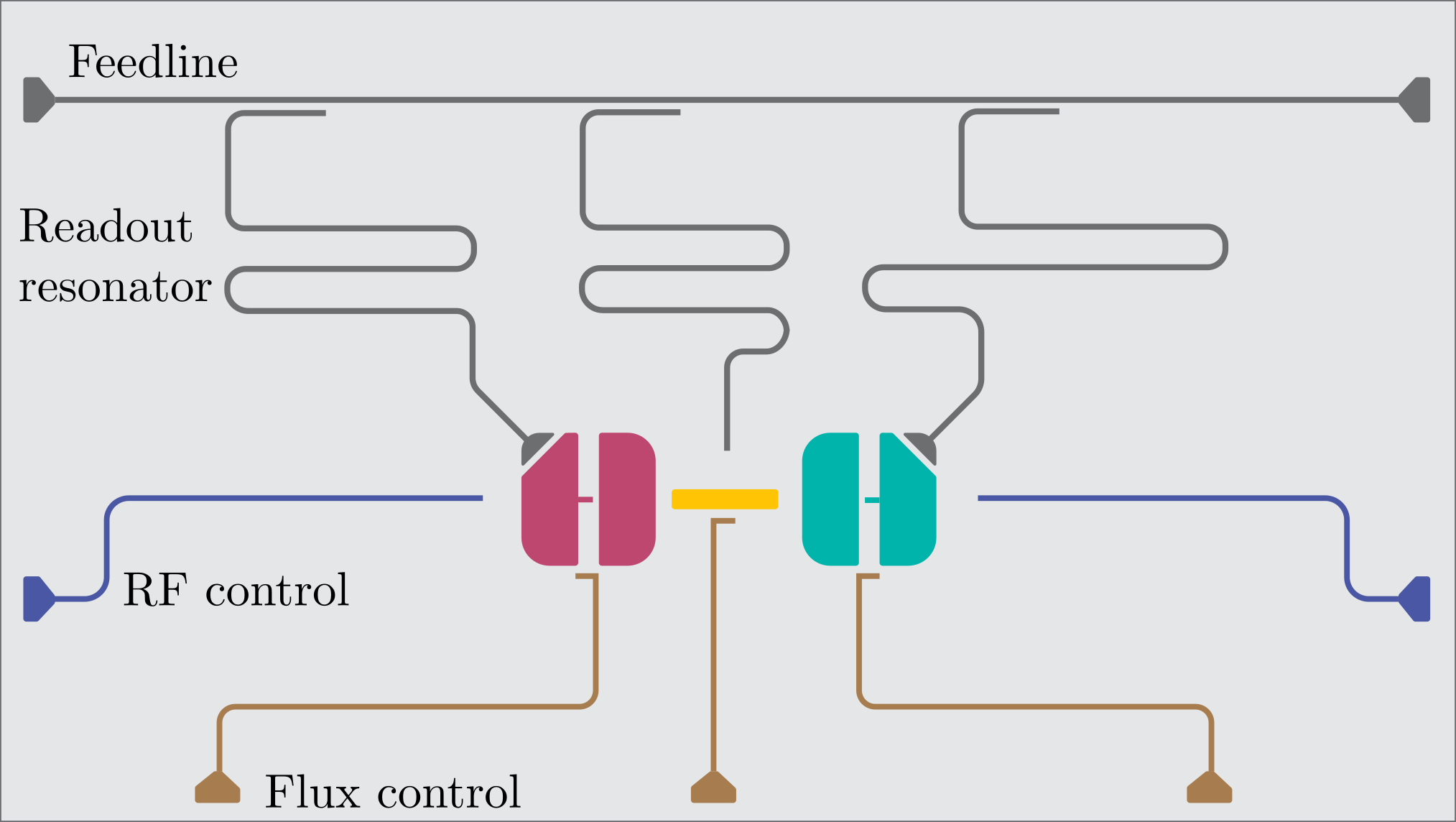}
    \caption{Schematic of the device used in the experiment. The floating qubits (fuchsia and teal) are coupled to rf (blue) and flux (brown) control lines while the tunable coupler (yellow) is connected to just a flux line. Both qubits and coupler are connected to individual readout resonators that in turn are coupled to a feed line (gray).}
    \label{fig:schematic}
\end{figure}
The parametric-resonance interaction Hamiltonian is obtained by setting $n=0$ in Eq.~\eqref{hamiltonian}\cite{Didier21}: \begin{align}\label{param-hamiltonian}
    H_{\rm int} &= \tilde g_{01} e^{-i\Delta t}|10\rangle
\langle 01|+\tilde g_{02}e^{-i(\Delta -\eta_2)t}|11\rangle\langle 02| \notag\\
&+ \tilde g_{20}e^{-i(\Delta +\eta_1)t}|20\rangle\langle 11| +\mathrm{H.c.},
\end{align}
with $\tilde g_{01}= gJ_{0}(\tilde \omega_2/2\omega_p)$, $\tilde g_{02}=\tilde g_{20} = \sqrt{2}gJ_{0}(\tilde \omega_2/2\omega_p)$. 

An iSWAP gate is activated when the modulated qubit 2 is brought into resonance with the other qubit for a half cycle of oscillation of population $|10\rangle$, whereas a CZ gate is enacted by bringing $|11\rangle$ and $|02\rangle$ (or $|20\rangle$) into resonance by tuning the average frequency of the modulated qubit for a full cycle of  population oscillation. 
Note that the activation of the parametric-resonance gate does not rely on the modulation frequency of the flux pulse $\omega_p$. However, in practice, to avoid sideband transitions and collisions, the modulation frequency is chosen such that $\omega_p$ is higher than any modulation frequency of sideband gates. Compared to sideband-activated parametric interactions, parametric-resonance interactions allow for faster entangling gates for the same bare coupling, $g$. This is because, for a small amplitude of modulation the parametric-resonance interaction has a renormalization factor, $J_0\approx 1$ with the sideband ($n\neq 0$) weights approaching zero. In comparison, for typical sideband gates the renormalization factor $(\mathrm{with} \ n \neq 0)$ is approximately $ 0.6$ \cite{Didier18}.

\begin{figure}[t]
    \centering
   \includegraphics[width=\linewidth]{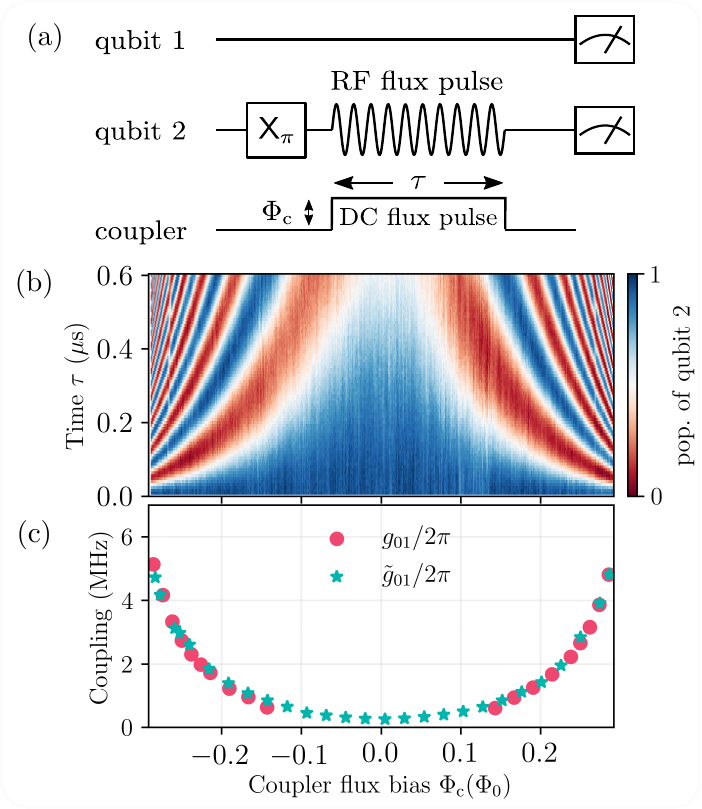}
    \caption{(a) Pulse sequence to measure qubit-qubit coupling under modulation as a function of the coupler flux bias. (b) Measured population of $|01\rangle$ versus pulse duration $\tau$ and coupler flux bias $\Phi_{\rm c}$. (c) Bare qubit-qubit coupling $g_{01}/2\pi$ (dotted fuchsia) (measured by replacing rf flux pulse in (a) by fast flux pulse) and effective qubit-qubit coupling $\tilde g_{01}/2\pi$ (star teal) versus coupler flux bias.}
    \label{fig:g01}
\end{figure}

\begin{figure*}[t]
    \centering
    \includegraphics[width=\linewidth]{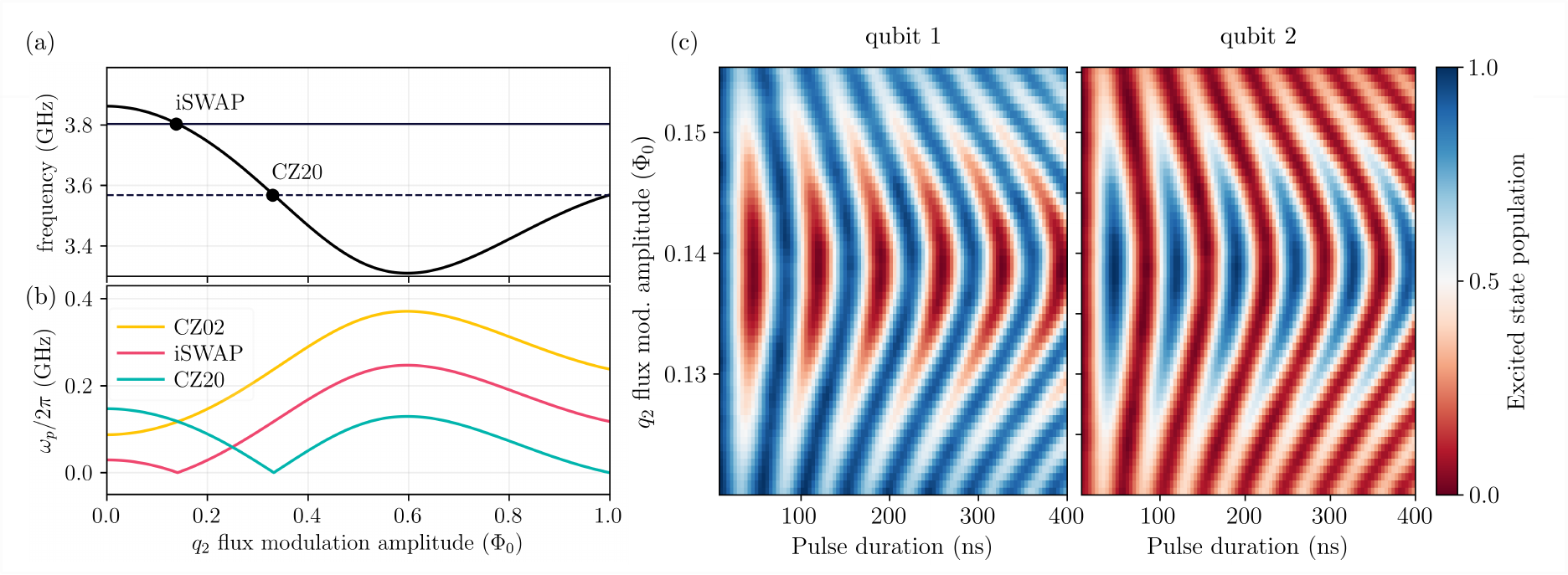}
    \caption{ (a) Average frequency of qubit 2 as a function of the flux modulation amplitude. The iSWAP gate is activated at resonance with $\omega_1/2\pi$ (solid line) and CZ20 at $(\omega_1-\eta_1)/2\pi$ (dashed line).
    (b) Qubit 2 flux modulation frequency $\omega_p$ versus flux modulation amplitude for $n=-2$ sideband [$\omega_{p, \rm iSWAP} = (\overline{\omega}_2-\omega_1)/2$, $\omega_{p, \rm CZ02} = (\overline{\omega}_2-\omega_1-\eta_2)/2$, $\omega_{p, \rm CZ20} = (\overline{\omega}_2-\omega_1+\eta_1)/2$].  The plots are generated using the measured parameters of the qubits and the coupler. (c) Measured energy exchange (chevron) between $|10\rangle$ and $|01\rangle$ for activating the parametric-resonance iSWAP interaction once a $\pi$ pulse has been applied to qubit 1.}
    \label{fig:qubit_chevron}
\end{figure*}

It is worth noting that, since most of the sideband weight is concentrated in the $n=0$ index when operating parametric-resonance gates, the effective coupling of sidebands $n \neq 0$ is strongly reduced. As a result, the sideband resonances are weakly coupled to two-level system (TLS) defects, potentially making parametric-resonance gates less susceptible to TLS.

\section{Parametric-resonance gate with tunable coupler} 
The interaction Hamiltonian in Eq.~\eqref{param-hamiltonian} describes a system of two capacitively coupled qubits. Parametric-resonance gates can also be implemented by coupling the qubits with a tunable coupler, a useful element to reduce the residual ZZ interaction while idling. In this work, we implement parametric-resonance gates using a grounded tunable coupler as shown in Fig.~\ref{fig:schematic}. The Hamiltonian in the interaction picture of this three-body system is given by (see Appendix \ref{Ham_derivation})
\begin{align}\label{Hqq}
H &= \sum_{k=1,2,c} (\omega_{k}\sigma_{01,k}^{\dag}\sigma_{01,k} + (2\omega_k-\eta_k)\sigma_{12,k}^{\dag}\sigma_{12,k})\notag\\
&\phantom{=}+ g_{1c}\sigma_{1\rm x}\sigma_{c\rm x}+g_{2c}\sigma_{c\rm x}\sigma_{2\rm x}+g_{12}\sigma_{1 \rm x}\sigma_{2\rm x},
\end{align}
where $\sigma_{k,01}=|0\rangle_k\langle 1|$, $\sigma_{k,12}=|1\rangle_k\langle 2|$, $\sigma_k = \sigma_{01,k} + \sqrt{2}\sigma_{12,k}$, $\sigma_{k\rm x} = \sigma_k^{\dag}+\sigma_k$; the $\omega_k$ are the $|0\rangle\rightarrow |1\rangle$ transition frequencies of qubit $k$ and coupler, $g_{jc}\ (j=1,2)$ is the qubit-coupler coupling strength, and $g_{12}$ is the direct qubit-qubit coupling. The couplings $g_{jc}$ depend on the flux bias applied to the tunable coupler. In particular, since the tunable coupler frequency is tuned by a few gigahertz, the coupling substantially changes with applied flux. The effective parametric-resonance interaction between the qubits (Appendix  \ref{Appendix_gqq_modulation}) has the same form as Eq.~\eqref{param-hamiltonian} but with new coupling strengths $\tilde g_{m}= J_0 (\tilde \omega_2/2\omega_p) g_{m}, m\in\{01,02,20\}$, where the bare coupling rates are
\begin{align}
    g_{\rm 01} &= g_{12} -\frac{g_{1c}g_{2c}}{2}\left(\frac{1}{\Delta_1} + \frac{1}{\Delta_2}\right),\\\label{Gqq}
     g_{\rm 02} &= \sqrt{2}g_{12} -\frac{\sqrt{2}g_{1c}g_{2c}}{2}\left(\frac{1}{\Delta_1} +\frac{1}{\Delta_2+\eta_2}\right),\\
      g_{\rm 20} &= \sqrt{2}g_{12} -\frac{\sqrt{2}g_{1c}g_{2c}}{2}\left( \frac{1}{\Delta_2}+\frac{1}{\Delta_1+\eta_1}\right),
\end{align}
with $\Delta_1 = \omega_c-\omega_1$ and $\Delta_2 = \omega_c-\overline{\omega}_2$. The coupling strength $\tilde g_{01}$ describes the exchange interaction rate between $|10\rangle$ and $|01\rangle$, whereas  $\tilde g_{02}$ and $\tilde g_{20}$ describe coupling strengths between $|11\rangle$ and $|02\rangle$, and between $|11\rangle$ and $|20\rangle$, respectively. In this system, the zero-coupling condition can be achieved by tuning the coupler frequency such that the virtual interaction mediated by the coupler $g_{1c}g_{2c}(1/\Delta_1+1/\Delta_2)/2$ offsets the direct qubit-qubit coupling $g_{12}$ when the coupler is above the qubit frequencies.

\section{Experimental results}\label{experiment}
The device is fabricated using standard lithographic techniques on a high resistivity Si wafer that is then diced into individual chips. The superconducting circuit components, including the qubit capacitor pads, Nb ground planes, signal routing, coplanar waveguides and readout resonators, are defined with an etching procedure \cite{Ani2019}. The Josephson junctions in the superconducting quantum interference device loops of the transmon qubits are then fabricated through double angle evaporation of Al.

The frequencies of the qubits ($\omega_1/2\pi = 3.173-3.803$ GHz, $\omega_2/2\pi = 3.207-3.862$ GHz) are tuned by applying an external magnetic flux through their SQUID loops. Single qubit X and Y rotations are performed with on-resonance 60 ns derivative removal by adiabatic gate \cite{Motzoi10} Gaussian-shaped microwave pulses applied through dedicated drive lines. The tunable coupler has symmetric junctions and has a maximum frequency of $\omega_c/2\pi = 5.915$ GHz. Qubit and tunable coupler frequencies are tuned with dedicated on-chip flux bias lines. Readout is performed with coplanar waveguide resonators inductively connected to a common feed line and operated in the dispersive regime (see the experimental setup in Appendix \ref{Experimental_setup}).

We first characterize the net qubit-qubit coupling strength $\tilde g_{01}$ by measuring the energy exchange between $|10\rangle$ and $|01\rangle$ as a function of the coupler flux bias. Both the qubits and the coupler are parked at their maximum frequencies while idling. To measure the energy exchange, we prepare $|01\rangle$ by applying a $\pi$ pulse followed by a modulated flux pulse $\Phi_{e2}$ ($\omega_p/2\pi = 300\ \mathrm{MHz}$) on qubit 2 to bring the two qubits into resonance for a variable pulse duration $\tau$ [see Fig~\ref{fig:g01}(a)]. During the parametric modulation, the qubit-qubit coupling strength is varied by tuning the coupler frequency with a fast flux pulse, $\Phi_{ec}(\tau) = \Phi_{\rm dc,c}+\Phi_{\rm c}u(\tau)$, where $\Phi_{\rm c}$ is the amplitude of the flux pulse and $u(\tau)$ is the pulse envelope. The populations of the two qubits are measured as a function of the rf flux pulse duration $\tau$ and the amplitude $\Phi_{c}$ of the dc flux pulse applied to the coupler [Fig.~\ref{fig:g01}(b)].

The period of oscillation increases gradually as the coupler flux bias decreases in magnitude. To quantify the qubit-qubit coupling strength $\tilde g_{01}$, we fit the population oscillation of $|01\rangle$ at every coupler flux bias [Fig.~\ref{fig:g01}(c)]. We also measured the bare coupling $g_{01}$ and found similar values, confirming the theory that predicts a renormalization factor close to 1 ($J_0 = 0.998$)~[see Appendix \ref{Device}]. A minimum coupling $\tilde g_{01}/2\pi\approx 250 ~\mathrm{kHz}$ occurs at zero flux bias and stays relatively constant for a wide range of flux bias (here we removed a hardware related offset). Note that these measurements are run after correcting substantial dc flux crosstalk; see Appendix \ref{flux_crosstalk} for details.

The residual ZZ coupling, when the qubits are parked at the idle bias, does not necessarily vanish at the same coupler flux bias where $\tilde g_{01}\approx 0$. To verify that the static ZZ is minimal at the idling point, we run single-qubit simultaneous randomized benchmarking (RB) measurements on the qubits and check that the fidelity is not severely impacted. We measured RB fidelities of $99.84 \pm 0.02 \%$ and $99.91 \pm 0.01\%$ for qubits 1 and 2, respectively. Compared to the nonsimultaneous RB $99.94 \pm 0.02\%$ (qubit 1) and $99.95 \pm 0.02\%$ (qubit 2), the corresponding decrease in fidelities are $0.1\%$ and $0.04\%$, which could be attributed to small but finite static ZZ coupling.

To identify the parametric-resonance gate operating points from measured qubit parameters, we calculate and plot the average frequency of the modulated qubit versus the modulation amplitude [Fig.~\ref{fig:qubit_chevron}(a)]. For the iSWAP gate, we choose a flux amplitude that gives $\overline{\omega}_2= \omega_1$ and for CZ20 we choose the amplitude that yields $\overline{\omega}_2=\omega_1-\eta_1$. Another way to determine the parameters for the resonance conditions is to plot the flux modulation frequency that activates the $n=-2$ sideband gates versus modulation amplitude [Fig.~\ref{fig:qubit_chevron}(b)]. The amplitudes that correspond to $\omega_p=0$ are the parametric-resonance gate operating points. Although these gates do not rely on a specific modulation frequency, the modulation frequency is chosen to be high enough so that collisions with the sideband-activating frequencies are avoided. For example, in Fig.~\ref{fig:qubit_chevron}(b), it is reasonable to choose $\omega_p/2\pi \gtrsim 280~\mathrm{MHz}$ for both iSWAP and CZ gates to avoid frequency collisions with sideband transitions. In practice, modulating the qubit at higher frequencies leads to significant attenuation of the signal coming from the hardware and fridge lines. This requires measuring the transfer function before choosing a modulation frequency; see Appendix \ref{transfer}.

Once the modulation frequency is identified, we select the coupler flux amplitude $\Phi_{\rm c}$ to target a qubit-qubit coupling while maintaining the qubit-coupler system in the dispersive regime. We first consider an iSWAP interaction and a coupling strength that yielded a 44 ns gate. A coupling strength larger than this resulted in leakage to the coupler, substantially impacting the gate fidelity. For a fixed $\Phi_{\rm c}$, we measure the energy exchange interaction between $|10\rangle$ and $|01\rangle$ by applying a $\pi$ pulse followed by a flux pulse modulated at $\omega_p/2\pi = 300~\mathrm{MHz}$ on qubit 2 to bring the qubits in resonance. Figure~\ref{fig:qubit_chevron}(c) shows the population transfer between qubits 1 and 2 as a function of the amplitude and duration of the modulated flux pulse. The amplitude and duration that gives full population transfer is used to activate the iSWAP gate. 


\begin{figure}
    \centering
    \includegraphics[width=\columnwidth]{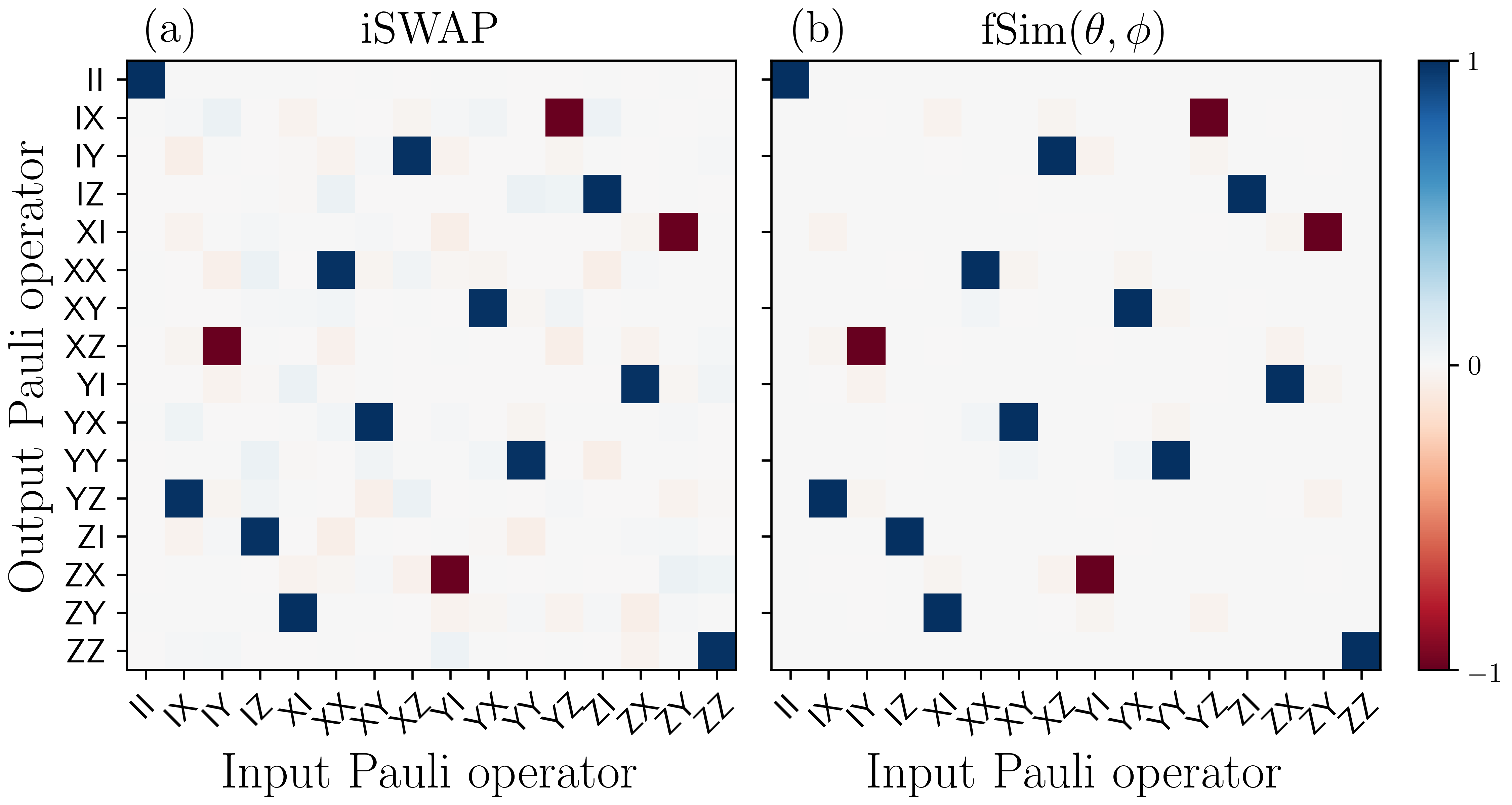}
    \caption{ (a) The iSWAP quantum process tomography with an iSWAP fidelity of $99.3\%$. (b) Reconstructed  quantum process tomography of $\mathrm{fSim}(\theta,\phi)$ at angles $\theta=-1.57$, $\phi= 0.076$, yielding a maximum possible fidelity against an arbitrary fSim of $99.4\%$.}
    \label{fig:iSWAP}
\end{figure}

\begin{figure}
    \centering
    \includegraphics[width=\columnwidth]{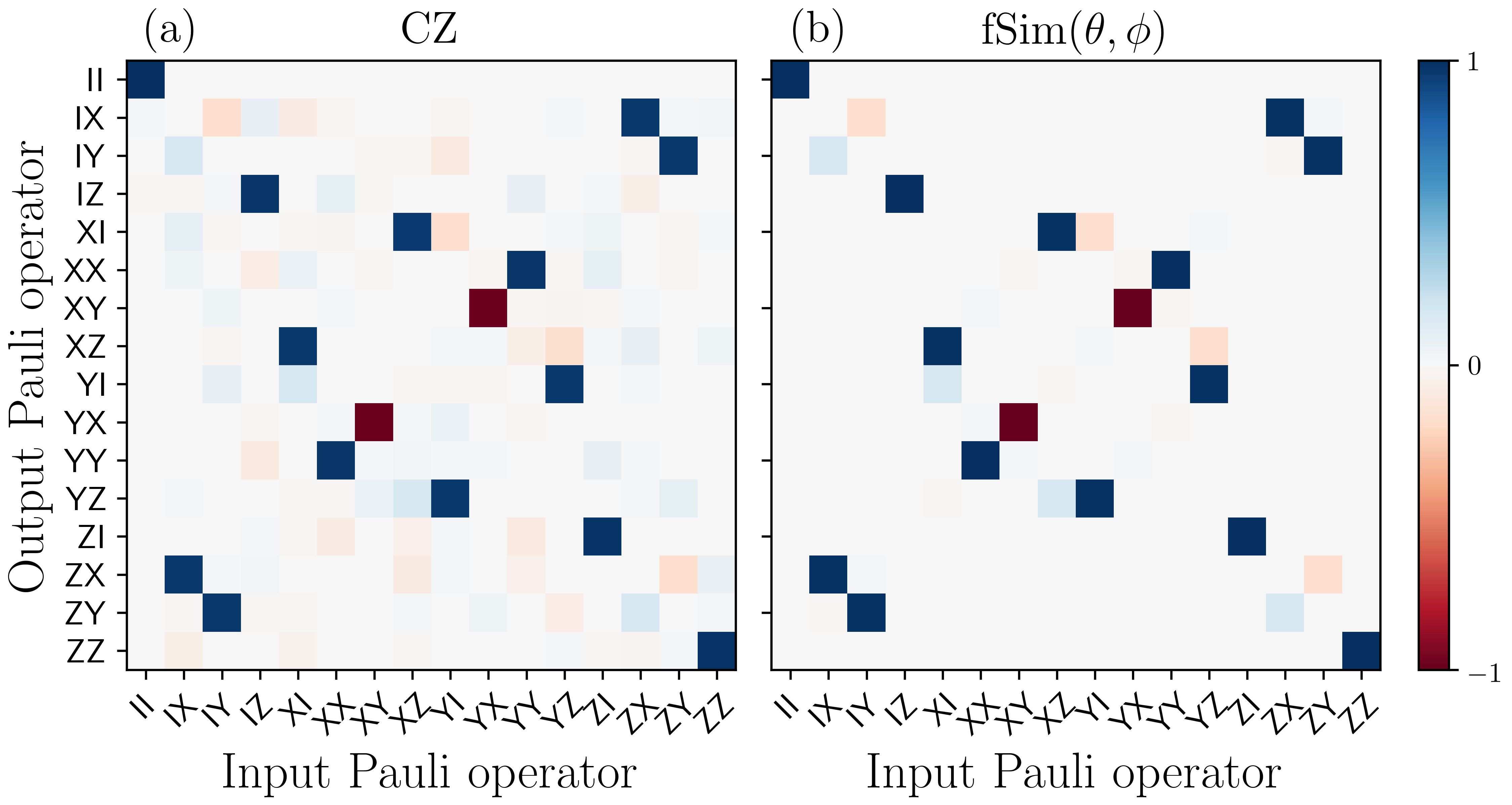}
    \caption{ (a) Quantum process tomography of a CZ gate with a fidelity of $97.9\%$. (b) Reconstructed  quantum process tomography of $\mathrm{fSim}(\theta,\phi)$ at angles $\theta=-0.006$, $\phi= 2.8$, yielding a maximum possible fidelity against an arbitrary fSim of $98.4\%$.}
    \label{fig:CZ}
\end{figure}

To assess the performance of the iSWAP gate, we use quantum process tomography (QPT) \cite{Hradil04,Horodecki99,Nielson02}. To estimate the QPT gate fidelities, we compensate for the readout errors \cite{Reagor18} and apply complete-positivity constraints. We estimated an average fidelity of $99.3\%$ for a gate with duration of $44~\mathrm{ns}$ [see Fig.\ \ref{fig:iSWAP}(a)]. At the operating point of the iSWAP interaction the relaxation and coherence times of both qubits are: $T_{1,q1}=70~\mu\mathrm{s}$, $T_{2,q1}^{*} = 14\ \mu\mathrm{s}$, $\tilde T_{1,q2} = 56\ \mu\mathrm{s}$, $\tilde T_{2,q2}^* = 10\ \mu\mathrm{s}$ (the tilde signifies a measurement made under modulation). The average coherence-limited fidelity is approximated by $\mathcal{F}\approx 1- (T_{1,q1}^{-1}+\tilde T_{1,q2}^{-1})\tau/5-2(T_{2,q1}^{* -1}+\tilde T_{2,q2}^{*-1})\tau/5$. Hence, the coherence-limited fidelity for $\tau = 44$ ns gate is $99.67\%$, indicating that the iSWAP average fidelity is predominantly limited by the qubits' dephasing times. The phase error due to the repulsion of higher energy levels can be estimated by comparing the measured process matrix with that of the native fSim tomography. The exchange interaction within the qubit subspace is a generator for the so-called fermionic simulation (fSim) gate unitary \cite{Babbush17} 
\begin{align}\label{fsim}
\mathrm{fSim}(\theta,\phi) = \left(
    \begin{array}{cccc}
         1  & 0 & 0  & 0\\
         0 & \cos \theta & -i\sin \theta & 0\\
         0 &-i\sin \theta & \cos\theta & 0\\
         0& 0  &0  & e^{-i\phi}
    \end{array}
    \right),
\end{align}
where $\theta$ is the swap angle and $\phi $ is the phase of $|11\rangle$ accumulated during the gate time due to interaction between $|11\rangle$ and $|02\rangle$, $|20\rangle$, and coupler states. Note that setting $\theta = -\pi/2 $ and $\phi = 0$ in Eq.~\eqref{fsim} yields the iSWAP unitary. Using the same data from iSWAP tomography, we estimate the fSim QPT fidelity to be $99.4\%$ with $\theta= -1.57$ and $\phi = 0.076$ [see Fig.~\ref{fig:iSWAP}(b)]. The phase error $3(1-\cos \phi )/10$ is calculated to be $0.09\%$, indicating that iSWAP fidelity is very close to that of the native fSim gate. A repeated iSWAP tomography over a span of 2 hours shows the gate fidelity varying from $96.2\%$ to $99.3\%$ (see Appendix \ref{stability}).

We also demonstrate a CZ gate by bringing the average frequency of qubit 2 in resonance with the $|1\rangle\rightarrow |2\rangle$ transition frequency of qubit 1, as illustrated in Fig.~\ref{fig:qubit_chevron}(a). A modulation frequency of $\omega_p/2\pi = 280\ \mathrm{MHz}$ is chosen to avoid frequency collisions with the sideband resonances. In order to minimize coherent leakage to the coupler, the coupler is biased such that the qubit-coupler system remains in the dispersive regime. To this end, biasing the coupler at $\Phi_{\rm c} = -0.29\Phi_0$, the gate time is set at $\tau= 124$ ns. The flux modulation amplitude of qubit 2 is chosen to maximize the population of $|11\rangle$ after full oscillation with $|20\rangle$. We measured a QPT fidelity of $97.9\%$ for the CZ gate and $98.4\%$ for the CZ-like fSim gate with $\theta = -0.006$ and $\phi = 2.8$ rad [see Fig.~\ref{fig:CZ}]. The coherence-limited fidelity estimated using $\mathcal{F} \approx 1-19(T_{1,q1}^{-1}+\tilde T_{1,q2}^{-1})\tau/160-(61T_{2,q1}^{* -1}/80 + 29\tilde T_{2,q2}^{*-1}/80)\tau$ \cite{Schuyler} and measured coherence times at the gate operating point is $98.83\%$, indicating  that CZ fidelity is not limited by qubit coherence. We estimate the phase error using $\delta \phi = \pi-2.8$ and $3(1-\cos \delta\phi)/10$ to be $1.73\%$, implying significant phase error due to the repulsion from higher levels of the qubits and coupler \cite{Oliver20}. Note that, since the CZ gate is longer than the iSWAP gate, it acquires more phase error.

To reduce the impact of flux noise under modulation, parametric gates are typically operated at the ac sweet spot~\cite{Didier20,Hong20,Schuyler}.
Because the modulation amplitude of the qubit is determined by the detuning between the qubits, parametric-resonance gates may not necessarily be operated at the ac sweet spot, resulting in short dephasing times. To mitigate the impact of flux noise, one can design qubits close in frequency such that the modulation only steers the qubit by tens of megahertz from the dc sweet spot. Alternatively, one can create a dynamical sweet spot at the gate operating flux bias of the qubit by using bichromatic modulation \cite{Didier21,Koch20}, which can also be used to optimize the sideband weights. 

\section{Conclusion}
We have demonstrated parametric-resonance entangling gates in a tunable coupler architecture. By bringing the qubits into resonance via parametric flux modulation, we approximately recover the bare coupling rate that is otherwise reduced in sideband activated parametric gates \cite{Caldwell18,Reagor18}. We have realized high-fidelity iSWAP and CZ gates on the same edge, potentially enabling more efficient decompositions for quantum circuits that require large numbers of swap operations~\cite{Abrams20}. Benchmarking parametric-resonance gates with other metrics such as randomize benchmarking is a subject of further investigation.

\begin{acknowledgements}
We thank Joseph Valery for insightful discussion on phase calibration and initial exploration of bichromatic modulation. We thank Joseph Valery, Alex Hill, Gregory Stiehl, and Matthew Reagor for critical reading of the manuscript. We also thank Glenn Jones and Douglas Zorn for developing the control hardware used in the experiments.
\end{acknowledgements}

\section*{Contributions}
E.A.S. wrote the manuscript, performed the experiments, and analyzed data. E.A.S. developed the theory of floating tunable qubits coupled via a grounded coupler, and the qubit-qubit coupling rates. N.D. suggested the idea of running parametric-resonance gates and derived the qubit-qubit coupling under modulation. A.Q.C and S.K. provided support for measurement tools. R.M. simulated and designed the device. S.P. developed the flux crosstalk measurement. All authors contributed to the editing of the manuscript and figures. S.P. coordinated the effort.

\appendix

\begin{widetext}
\section{Derivation of the effective qubit-qubit Hamiltonian} \label{Ham_derivation}
We consider two ``floating'' tunable transmon qubits (fuchsia and teal) capacitively coupled to a ``grounded'' tunable transmon coupler (orange) [see Fig. \ref{fig:circuit}]. To derive the Hamiltonian of the coupled system, we first write the Lagrangian in terms of the node fluxes and the circuit elements following the methods in Refs.~\cite{Devoret97,Girvin2014} 
\begin{align}
    \mathcal{L} & = T-U\\
    T &= \frac{1}{2}C_{01}\dot\Phi_1^2+\frac{1}{2}C_{02}\dot\Phi_2^2+\frac{1}{2}C_{12}(\dot\Phi_2-\dot\Phi_1)^2 + \frac{1}{2}C_{23}(\dot\Phi_3-\dot\Phi_2)^2+\frac{1}{2}C_{13}(\dot\Phi_3-\dot\Phi_1)^2\notag\\
    &+\frac{1}{2}C_{03}\dot\Phi_3^2 + \frac{1}{2}C_{43}(\dot\Phi_4-\dot\Phi_3)^2 + \frac{1}{2}C_{35}(\dot\Phi_5-\dot\Phi_3)^2+\frac{1}{2}C_{45}(\dot\Phi_5-\dot\Phi_4)^2+\frac{1}{2}C_{04}\dot\Phi_4^2 + \frac{1}{2}C_{05}\dot\Phi_5^2 + \frac{1}{2}C_{24}(\dot\Phi_4-\dot\Phi_2)^2\notag\\
    U & = -E_{J1}\cos(\phi_{2}-\phi_{1}+\phi_{01})-E_{Jc}\cos(\phi_{0c}+\phi_{3})-E_{J2}\cos(\phi_{4}-\phi_{5}+\phi_{02}),
\end{align}
where 
\begin{align}\label{EJs}
    E_{J} (\phi_{ek})& = \sqrt{E_{JSk}^2+E_{JLk}^2+2E_{JSk}E_{JLk}\cos(\phi_{ek})},\\
    \phi_{0k} &= \tan^{-1}\left[\frac{E_{JSk}-E_{JLk}}{E_{JSk}+E_{JLk}} \tan(\phi_{ek}/2)\right]\label{phi0}, ~~~k \in \{1,2,c\}
\end{align}
and $E_{Jk} = E_{JLk}+E_{JSk}$ with $E_{JLk}$ and $E_{JSk}$ being the Josephson energies for the large and small junctions of the SQUIDs and the $C_{ij}$ are capacitances; $\phi_{l} = 2\pi \Phi_{l}/\Phi_{0}$, ($l=1,\hdots,5$) with $\Phi_0= h/2e$ the flux-quantum and $\Phi_{l}$ node fluxes, $\phi_{ek} = 2\pi \Phi_{ek}/\Phi_0$ is the external flux bias through the loops. The node fluxes are defined in terms of node voltages as $\Phi_{l}= \int_{-\infty}^{t}dt'V_l(t')$, where $V_l$ is the voltage at the $j$th node (represented by the solid circles in the circuit). Introducing new variables $\Phi_{1p/m}= \Phi_{2}\pm \Phi_{1} $, $\Phi_{2p/m} = \Phi_{4}\pm\Phi_{5}$  and $\Phi_{c} \equiv \Phi_3$, 
the canonical momenta $Q_k$ and coordinates $\Phi_k$, related as $Q_k = \partial \mathcal{L}/\partial \dot \Phi_{k}$, can be written in matrix form as $\mathbf{Q} =\mathbf{C}\dot{\mathbf{\Phi}}$, i.e.,: 
\begin{equation}
\left(
\begin{array}{c}
     Q_{1p}  \\
     Q_{1m} \\
     Q_{c}\\
     Q_{2p}\\
     Q_{2m}
\end{array}
\right)= \frac{1}{4}
\left(
\begin{array}{ccccc}
     C_{1p}& C_{1m} &-2(C_{13}+C_{23})&-C_{24}& -C_{24}\\
     C_{1m}& C_{1p} +4C_{12}& -2(C_{23}-C_{13})& -C_{24} &-C_{24}\\
     -2(C_{13}+C_{23})&-2(C_{23}-C_{13})&4C_{cp}&-2(C_{34}+C_{35})&-2(C_{34}-C_{35})\\
     -C_{24}& -C_{24}& -2(C_{34}+C_{35})&C_{2p}&C_{2m}\\
     -C_{24}&C_{24}&-2(C_{34}-C_{35})&C_{2m}&C_{2p}+4C_{45}
\end{array}
\right)\left(
\begin{array}{c}
     \dot \Phi_{1p}  \\
     \dot \Phi_{1m} \\
     \dot \Phi_c\\
     \dot \Phi_{2p}\\
     \dot \Phi_{2m}
\end{array}
\right),
\end{equation}
where
\begin{align}
    C_{1p/m} &= C_{02}+C_{23}+C_{24}\pm (C_{01}+C_{13}),\\
    C_{2p/m} &= C_{04}+C_{34}+C_{24}\pm (C_{05}+C_{35}),\\
    C_{cp} &= C_{03}+C_{13}+C_{23}+C_{34}+C_{35}.
\end{align}

\begin{figure*}
    \centering
    \includegraphics[width=0.5\linewidth]{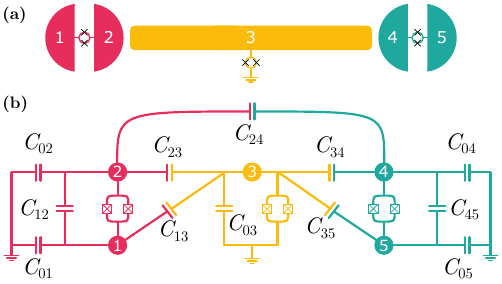}
    \caption{(a) Schematics of two floating transmon qubits (fuchsia and teal) coupled to a grounded tunable coupler (yellow).
    (b) Lumped-element circuit representation of the qubit-coupler-qubit system with the numbers showing the nodes that define the node fluxes used to derive the Hamiltonian.}
    \label{fig:circuit}
\end{figure*}
Note that the qubit modes are represented by flux variables $\Phi_{1m}$ and $\Phi_{2m}$ and the tunable coupler mode by $\Phi_{c}$. The modes represented by flux variables $\Phi_{1p}$ and $\Phi_{2p}$ are ``free particles'' rather than harmonic oscillators because their spring constant vanishes (there is no inductance associated with these modes). The Hamiltonian of the system is then given by $H=\frac{1}{2}\mathbf{Q}^{\rm T}\mathbf{C}^{-1}\mathbf{Q}+U$, 
\begin{align}\label{Ham-quant}
H &= 4E_{C1}\hat n_1^2+4E_{C2}\hat n_2^2+4E_{Cc}\hat n_c^2+4E_{1c}\hat n_1 \hat n_c+4E_{2c}\hat n_2 \hat n_c+4E_{12}\hat n_1\hat n_2\notag\\
&-E_{J1}\cos(\hat \phi_{1}+\phi_{01})-E_{Jc}\cos(\hat\phi_c+\phi_{0c})-E_{J2}\cos(\hat \phi_{2}+\phi_{02}),
\end{align}
where $\hat n_k=-i\partial /\partial \phi_k$ is the Cooper-pair number operator and $E_{Ck}=e^2/(2C_{\Sigma k})$ is the charging energy, where $C_{\Sigma c} = C_{cp}$. The qubit-coupler $C_{23},C_{34},C_{13},C_{35}$ and the qubit-qubit $C_{24}$ coupling capacitances are smaller than the capacitances between the qubit pads and the ground. We also note that the direct qubit-qubit coupling capacitance $C_{24}$ is smaller than the qubit-coupler coupling capacitances $C_{23}$ and $C_{45}$. Besides, the capacitance between the pads of the qubits is smaller than their capacitance to the ground, $\{C_{12},C_{34}\}< \{C_{01},C_{02},C_{04},C_{05}\}$. In reality the capacitances of the qubit pads to the ground are approximately equal for all qubits, i.e., $C_{01}\approx C_{02}$ and $C_{04}\approx C_{05}$. With these relationships and approximations, and assuming that $C_{13}= C_{35} =0$, the qubit-coupler coupling energies $E_{jc}$, and the direct qubit-qubit coupling energy $E_{12}$ are given by

\begin{align}
    E_{1c} &\approx \frac{e^2C_{23}}{2C_{cp}C_{\Sigma 1}},~~~ E_{2c} \approx \frac{e^2C_{34}}{2C_{cp}C_{\Sigma 2}},\\
    E_{12} &\approx \frac{e^2}{4C_{\Sigma 1}C_{\Sigma 2}C_{cp}}(C_{23}C_{34}+C_{24}C_{cp}). \label{qq-coupling}
\end{align}
The direct qubit-qubit coupling energy [Eq.~\eqref{qq-coupling}] has two terms. The first term describes the coupling mediated by the tunable coupler, while the second term is due to the direct capacitive coupling between the pads of the two qubits. Since $C_{cp}\gg C_{23},C_{34}$, the second term can have significant contribution even if $C_{24}$ is small compared to $C_{23},C_{34}$. The capacitance $C_{24}$ plays an important role in achieving a vanishing qubit-qubit coupling.\\

\noindent
{\textit{Harmonic oscillator basis}}. We introduce annihilation and creation operators for a harmonic oscillator as
\begin{align}
&\hat n_{k} = in_{k}^{\rm zpf}\left(a_k^{\dag}-a_k\right),~~~~\hat \varphi_{k} = \phi_{k}^{\rm zpf}\left(a_k^{\dag}+a_k\right), ~~k\in \{1,2,c\}.
\end{align}
where the zero-point fluctuations are given by
\begin{align}
    n_{k}^{\rm zpf} &= \frac{1}{\sqrt{2}}\left(\frac{E_{Jk}}{8E_{Ck}}\right)^{\frac{1}{4}},~~~~
    \varphi_{k}^{\rm zpf} =\frac{1}{\sqrt{2}}\left(\frac{8E_{Ck}}{E_{Jk}}\right)^{\frac{1}{4}},
\end{align}
where $[a_k,a^{\dag}_k]=1$. Subsisting these into Eq.~\eqref{Ham-quant} and expanding the cosine terms up to the fourth order, we get   
\begin{align}
    H & = \sum_{k=1,2,c}\left[\omega_{k}+\frac{E_{C_k}}{2}(1+\xi_k/4) -\frac{E_{C_k}}{2}(1+9\xi_k/16)a_k^{\dag}a_k\right]a_k^{\dag}a_k\notag\\
    &+ \sum_{k=1}^2g_{kc}\left(a_{k}a_{c}^{\dag}+a_{k}^{\dag} a_{c}-a_{k}a_{c}-a_{k}^{\dag} a_{c}^{\dag}\right)
    + g_{12}\left(a_{1}a_{2}^{\dag}+a_{1}^{\dag} a_{2}-a_{1}a_{2}-a_{1}^{\dag} a_{2}^{\dag}\right),
\end{align}
where the frequencies of the qubits and the tunable coupler are
\begin{align}
    \omega_{k} & = \sqrt{8E_{Jk}E_{Ck}}-E_{Ck} (1+\frac{\xi_k}{4}), ~~~\xi_{k} = \sqrt{2E_{Ck}/E_{Jk}}, k \in \{1,2,c\}
\end{align}
and the coupling strengths are
\begin{align}
    g_{jc} &= \frac{E_{jc}}{\sqrt{2}}\left(\frac{E_{Jj}}{ E_{Cj}} \frac{E_{Jc}(\Phi_{ec})}{E_{Cc}}\right)^{\frac{1}{4}} [1-\frac{1}{8}(\xi_{c}+\xi_{j})],~~j\in\{1,2\},\\
    g_{12} &= \frac{E_{12}}{\sqrt{2}}\left(\frac{E_{J1}}{ E_{C1}} \frac{E_{J2}}{E_{C2}}\right)^{\frac{1}{4}} [1-\frac{1}{8}(\xi_{1}+\xi_{2})]\label{q-q-g},
\end{align}
where the $g_{jc}$ are the qubit-coupler couplings and $g_{12}$ is the direct qubit-qubit capacitive coupling. The qubit-coupler couplings are directly proportional to the coupling energies  $E_{jc}$, which in turn are proportional to $C_{23}$ or $C_{34}$. Thus, these couplings can be controlled by varying $C_{23}$ or $C_{34}$. The correction terms $1 -(\xi_{c}+\xi_{j})/8$ are due to the qubits' nonlinearities. 

Approximating the qubits and the coupler by their lowest three energy levels, the Hamiltonian can be written as
\begin{align}
H& = H_0+ H_I +H_{12},\label{Hqq}\\
H_0& = \sum_{k=1,2,c} \omega_{k}\sigma_{k,01}^{\dag}\sigma_{k, 01} + (2\omega_k-\eta_k)\sigma_{k,12}^{\dag}\sigma_{k,12},\\ 
H_I & =  g_{1c}\sigma_{1\rm x}\sigma_{c\rm x}+g_{2c}\sigma_{c\rm x}\sigma_{2\rm x}+g_{12}\sigma_{1 \rm x}\sigma_{2\rm x},
\end{align}
where $\sigma_{k,01}=|0\rangle_k\langle 1|$, $\sigma_{k,12}=|1\rangle_k\langle 2|$, $\sigma_k = \sigma_{01,k} + \sqrt{2}\sigma_{12,k}$, $\sigma_{k\rm x} = \sigma_k^{\dag}+\sigma_k$;  $\omega_k$ and $\eta_k$ are the $|0\rangle\rightarrow |1\rangle$ transition frequency and absolute value of the anharmonicity of the qubits and coupler. Here, without loss of generality we used $\sigma_{\rm x}\sigma_{\rm x}$ coupling as opposed to $\sigma_{\rm y}\sigma_{\rm y} $ coupling. The effective qubit-qubit Hamiltonian can be derived by adiabatically eliminating the tunable coupler in the dispersive limit. We can approximately diagonalize the Hamiltonian applying the Schrieffer-Wolff transformation. Using the unitary operator defined by $U=e^{-S}$, the transformed Hamiltonian has the form 
\begin{align}
    H_{\rm eff} &= e^{-S}(H_0+H_I) e^{S} = H_0+H_I +[H_0+H_I,S]+\frac{1}{2}[[H_0,S],S]+\hdots,
\end{align}
where the generator $S$ satisfies $S=-S^{\dag}$.  This Hamiltonian can be made diagonal to the first order in $H_I$ by choosing the generator $S$ such that $[H_0,S] =- H_{I}$. Under this condition, the diagonalized Hamiltonian up to the second order reduces to
\begin{align}\label{Ham}
    H_{\rm eff}= H_{0} +\frac{1}{2}[H_I,S].
\end{align}
Evaluating the commutation $[H_I,S]$ and assuming that the coupler remains in the grounded state $\langle \sigma_{zc}\rangle = -1$, we derived the effective qubit-qubit Hamiltonian:
\begin{align}
    H_{qq}  &= \sum_{j=1,2}(\omega_{01,j} \sigma_{j,01}^{\dag}\sigma_{j, 01} + \omega_{02,j}\sigma_{j,12}^{\dag}\sigma_{j, 12}) \notag\\
    &+ g_{01}(|10\rangle\langle 01| + |01\rangle\langle 10|) + g_{02} ( |11\rangle \langle 02|+ |02\rangle\langle 11|)
    + g_{20} ( |11\rangle \langle 20| + |20\rangle \langle 11|),
\label{Hqqnomod}
\end{align}
where the dressed qubit frequencies and coupling between different transition are
\begin{align}
\omega_{01,k} &= \omega_k  + \frac{g_{kc}^2}{\Delta_k} + \frac{g_{kc}^2}{\Sigma_k},~~\omega_{02,k} = 2\omega_k-\eta_k  + \frac{2g_{kc}^2}{\Delta_k+\eta_k} + \frac{2g_{kc}^2}{\Sigma_k-\eta_k}, k\in \{1,2\},\\
    g_{01} &= g_{12} -\frac{g_{1c}g_{2c}}{2}\sum_{k=1,2}\left(\frac{1}{\Delta_k} + \frac{1}{\Sigma_k}\right),\label{gqq}\\
     g_{\rm 02} &= \sqrt{2}g_{12} -\frac{g_{1c}g_{2c}}{\sqrt{2}}\left(\frac{1}{\Delta_1} +\frac{1}{\Delta_2+\eta_2}+ \frac{1}{\Sigma_1}+\frac{1}{\Sigma_2-\eta_2}\right),\\
      g_{\rm 20} &= \sqrt{2}g_{12} -\frac{g_{1c}g_{2c}}{\sqrt{2}}\left(\frac{1}{\Delta_1+\eta_1} +\frac{1}{\Delta_2}+ \frac{1}{\Sigma_1-\eta_1}+\frac{1}{\Sigma_2}\right).
\end{align}
Note that the qubit-coupler coupling $g_{jc}(\Phi_{ec})$ depends on the coupler flux bias. Because of the large tunability of the tunable coupler, the couplings can be quite different at the extrema of the coupler's frequency.

\section{Qubit-qubit coupling under modulation}
\label{Appendix_gqq_modulation}

The Hamiltonian of two transmon qubits coupled through a direct coupling and via a tunable transmon is given in Eq.~\eqref{Hqq}.
In the following, we neglect the renormalization of the Fock states by the transmon nonlinearity as well as the weak dependence of the anharmonicity against flux.
A flux modulation, at frequency $\omega_p$, is now applied on qubit~2.
The transmon frequency as well of the couplings being flux dependent, $\omega_2(t)$, $g_{12}(t)$, $g_{2c}(t)$ become time dependent.
To derive the effective coupling between the transmons under flux modulation, we express the Hamiltonian in the interaction picture.
We then use the Fourier expansion
\begin{align}
g(t)e^{-i\int_0^t\mathrm{d}t'\omega(t')} &= \sum_ng_ne^{-i(\overline{\omega}+n\omega_p)t}, &
g_{12,n} &= \varepsilon_n g_{12}, &
g_{2c,n} &= \varepsilon_n g_{2c},
\end{align}
to express the Hamiltonian as a series,
\begin{align}
\widetilde{H}(t) &= \sum_n\widetilde{H}_n(t), &
\widetilde{H}_n(t) &= g_{12,n}X_1(t)X_{2,n}(t) + g_{1c}X_1(t)X_c(t) + g_{2c,n}X_{2,n}(t)X_c(t).
\end{align}
The renormalization coefficient $\varepsilon_n$ depends on the electrical parameters of the modulated transmon and the parameters of the flux modulation. Without loss of generality, we assume $\varepsilon_n$ to be real for simplicity of the present derivation. The time-dependent charge number operators are,
\begin{align}
X_n(t) &= \ket{0}\bra{1}e^{-i(\overline{\omega}+n\omega_p)t} + \sqrt{2}\ket{1}\bra{2}e^{-i(\overline{\omega}-\eta+n\omega_p)t} + \mathrm{H.c.}
\end{align}
with $X_{j\in\{1,c\},n}(t)\equiv X_{j,n=0}(t)$.

We consider the coupler in the dispersive regime, obtained when the detuning between the coupler frequency and the qubit frequencies and sidebands is much larger than their coupling strengths.
We then eliminate the tunable coupler using a Magnus expansion of the Hamiltonian in the interaction picture.
The effective Hamiltonian at second order is
\begin{align}
\widetilde{H}_\mathrm{eff}(t) &= \widetilde{H}(t) - \frac{i}{2}\int_0^t\mathrm{d}t'[\widetilde{H}(t),\widetilde{H}(t')].
\end{align}
We keep the terms that leave the tunable coupler in its ground state and generate a direct coupling between the qubits.
After eliminating the tunable coupler, we obtain,
\begin{align}
\widetilde{H}_\mathrm{eff}(t) &= \sum_n \widetilde{H}_{\mathrm{eff},n}(t),\\
\widetilde{H}_{\mathrm{eff},n}(t) &= g_{12,n}X_1(t)X_{2,n}(t) - g_{1c}g_{2c,n} \int_0^t\mathrm{d}t' [X_1(t)X_{2,n}(t')+X_1(t')X_{2,n}(t)]\sin[\omega_c(t-t')].
\end{align}
The time integral is then computed to provide the effective qubit-qubit coupling under modulation,
\begin{align}
\widetilde{H}_\mathrm{eff}(t) = \sum_n\left\{ g_{01,n}e^{-i(\Delta+n\omega_p)t}\ket{10}\bra{01} 
+ g_{02,n}e^{-i(\Delta-\eta_2+n\omega_p)t}\ket{11}\bra{02} 
+ g_{20,n}e^{-i(\Delta+\eta_1+n\omega_p)t}\ket{20}\bra{11} +\mathrm{H.c.} \right\}.
\end{align}
The effective couplings are given by
\begin{align}
g_{01,n} &= g_{12,n} -g_{1c}g_{2c,n} \frac{1}{2}\left( \frac{1}{\Delta_1} + \frac{1}{\Delta_2-n\omega_p} + \frac{1}{\Sigma_1} + \frac{1}{\Sigma_2+n\omega_p}\right),\\
g_{02,n} &= \sqrt{2}g_{12,n} - g_{1c}g_{2c,n}\frac{\sqrt{2}}{2}\left( \frac{1}{\Delta_1} + \frac{1}{\Delta_2+\eta_2-n\omega_p} + \frac{1}{\Sigma_1} + \frac{1}{\Sigma_2-\eta_2+n\omega_p} \right),\\
g_{20,n} &= \sqrt{2}g_{12,n} - g_{1c}g_{2c,n}\frac{\sqrt{2}}{2}\left( \frac{1}{\Delta_1+\eta_1} + \frac{1}{\Delta_2-n\omega_p} + \frac{1}{\Sigma_1-\eta_1} + \frac{1}{\Sigma_2+n\omega_p}\right),
\end{align}
with $\Delta_1=\omega_c-\omega_1$, $\Delta_2=\omega_c-\overline{\omega}_2$, $\Sigma_1=\omega_c+\omega_1$, $\Sigma_2=\omega_c+\overline{\omega}_2$.
The effective coupling thus depends on the average frequency of the modulated qubit and the modulation frequency.

Sideband gates can be implemented with a tunable coupler by compensating the detuning between the qubits with the modulation frequency. The parametric-resonance gates are obtained at the resonance condition for $n=0$. The effective coupling in the absence of flux modulation, Eq.~\eqref{Hqqnomod}, is recovered with $\overline{\omega}\equiv\omega$, $\varepsilon_{n=0}\equiv 1$, and $\varepsilon_{n\neq0}=0$.

For small modulation amplitudes around the maximum of the tunability band, the frequency under modulation oscillates at twice the modulation frequency around the average frequency $\overline{\omega}_2$ with an amplitude $\tilde \omega_2$. When the sideband $2n$ is used to parametrically activate a coherent exchange between the qubits, the renormalization factor is given by Bessel functions
\begin{align}
    \varepsilon_n \approx J_n(\tilde \omega_2/(2\omega_p)).
\end{align}
In particular, for the iSWAP gate implemented in this work, we have $\varepsilon_0\approx0.998$, $|\varepsilon_{\pm1}|\approx0.049$, and $|\varepsilon_{\pm2}|\approx0.001$.
\end{widetext}

\section{Experimental Setup}\label{Experimental_setup}
\begin{figure}
    \centering
    \includegraphics[width=\linewidth]{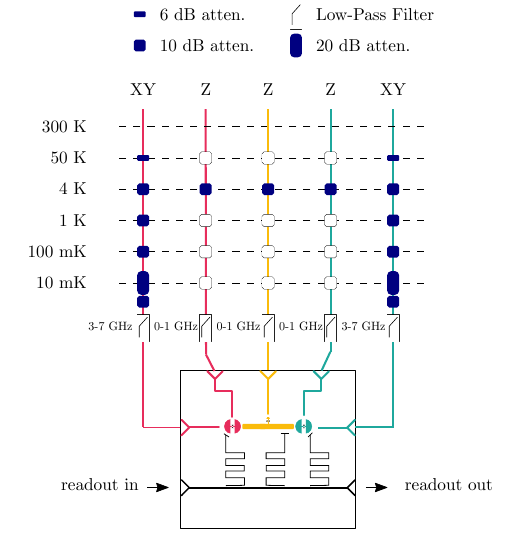}
    \caption{{\bf{Overview of the experimental setup}}. This diagram details the control electronics, wiring, and filtering for
the two qubits and the coupler. The readout signal, rf signal (XY lines), and dc and ac signals for flux delivery (Z lines) are generated by custom Rigetti arbitrary waveform generators. All control lines go through various stages of attenuation and filtering in the dilution refrigerator.}
    \label{fig:wiring}
\end{figure}

The physical device used in our experiment is packaged and mounted in a dilution refrigerator and cooled to 8 mK. The sample is mounted to a copper printed circuit board using Al wirebonds and packaged in a light-tight assembly. Direct current and microwave signals are delivered via non-magnetic sub miniature push-on surface mount connectors. An overview of the experimental setup used to measure the two qubits and the tunable coupler used in this experiment is shown in Fig.~\ref{fig:wiring} (individual components are labeled). 

\section{Device parameters}\label{Device}
\begin{figure}[t]
    \centering
    \includegraphics[width=\linewidth]{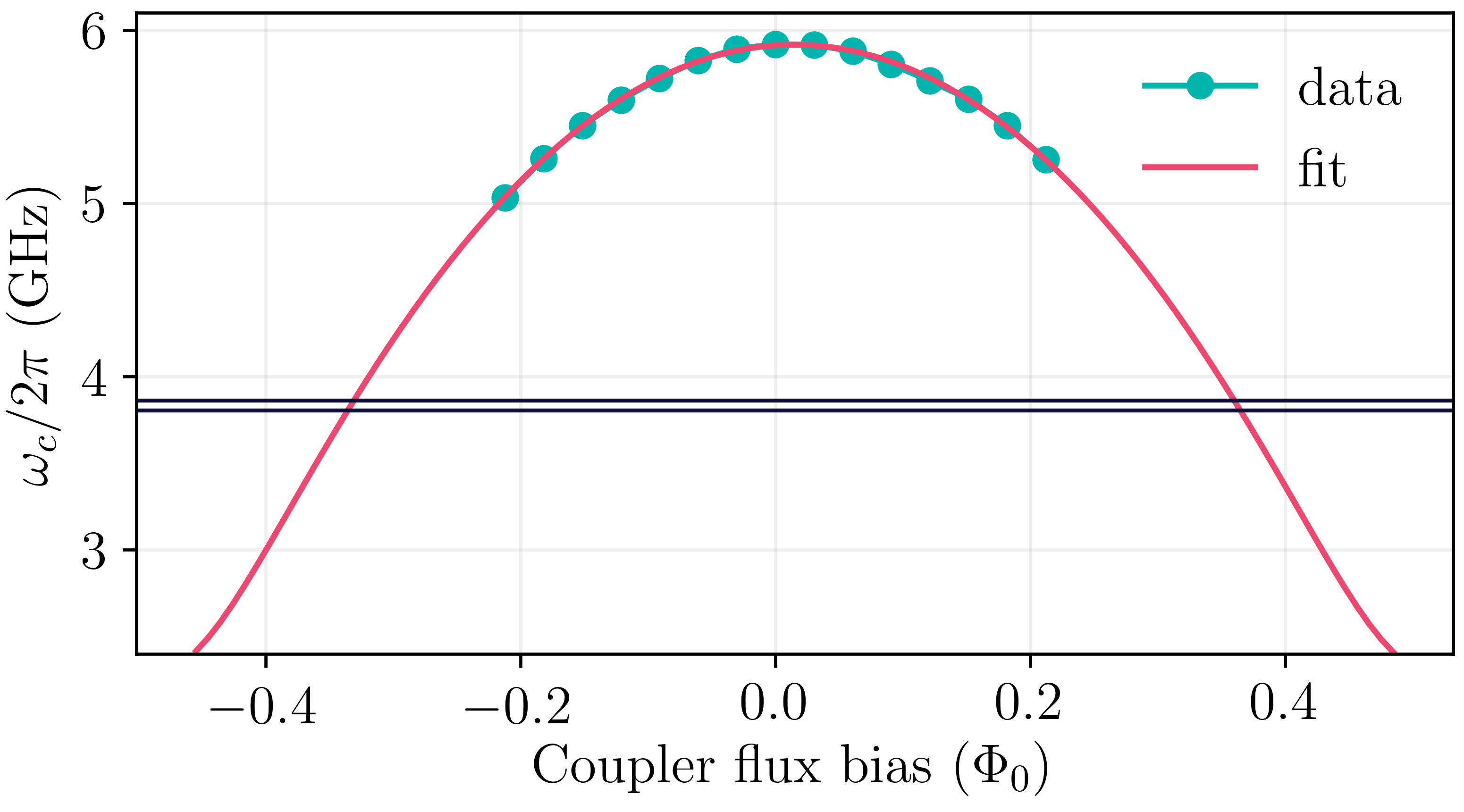}
    \caption{{\bf{Coupler frequency}}. Tunable coupler $|0\rangle\rightarrow|1\rangle$ transition frequency $\omega_c/2\pi$ versus tunable coupler fast flux amplitude $\Phi_{\rm c}$. The circles are measured data points and the solid fuchsia line is the fit to a transmon model. The black solid lines are the qubits' maximum frequencies.}
    \label{fig:coupler}
\end{figure}

\begin{figure}[tp]
    \centering
    \includegraphics[width=\linewidth]{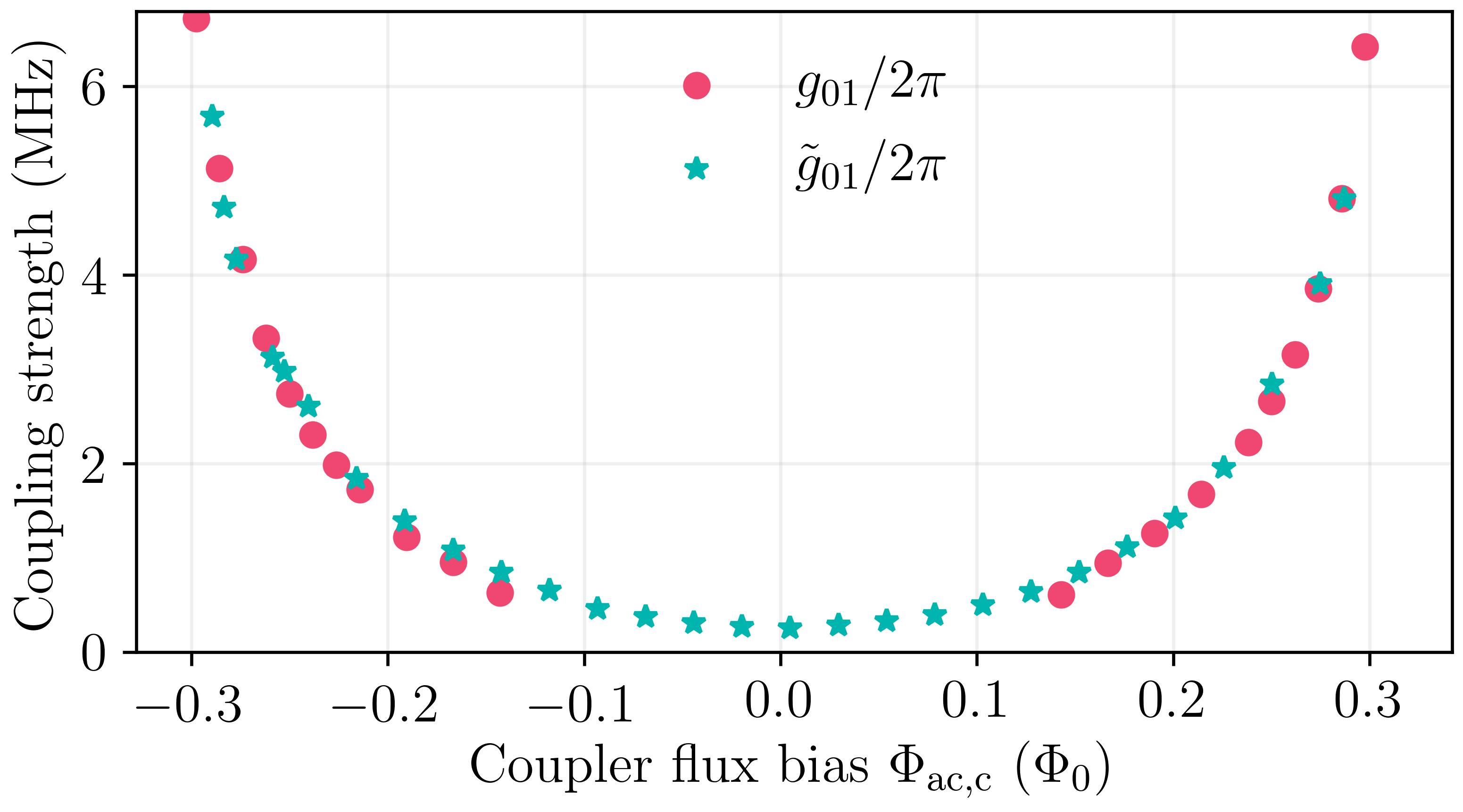}
    \caption{{\bf{Qubit-qubit couplings}.} Bare net qubit-qubit coupling strength $g_{01}/2\pi$ (dotted fuchsia) and renormalized net qubit-qubit coupling $\tilde g_{01}/2\pi$ (star teal) versus coupler flux bias. This shows that the measured bare and the renormalized coupling rates are the same within the accuracy of the measurement.}
    \label{fig:couplings}
\end{figure}

The device parameters are summarized in Table \ref{tab:parameters}. The coupler frequency versus flux bias and the idling frequencies of the qubits are shown in Fig.~\ref{fig:coupler}. The bare direct qubit-qubit coupling $g_{12}$ and qubit-coupler coupling $g_{jc}$ can be obtained by fitting the bare net qubit-qubit coupling $g_{01}$ versus coupler flux bias data to Eq.~\eqref{gqq}. The coupling $g_{01}$ is measured by preparing one of the qubits in the excited state, applying a fast flux pulse on the qubit and on the tunable coupler, and measuring either of the qubits. This measurement is repeated by varying the fast flux amplitude of the tunable coupler. We determine the coupling rate by fitting the population of $|10\rangle$ or $|01\rangle$ to a decaying cosine function for each amplitude of the coupler's flux bias. The effective qubit-qubit coupling $\tilde g_{01}$ under modulation is measured by repeating the above steps, but replacing the fast flux pulse on the coupler by a modulated flux pulse. Consistent with theory, coupling rates $\tilde g_{01}$ and $g_{01}$ are the same within the measurement accuracy because the renormalization factor is close to 1; see Fig.~\ref{fig:couplings}. When measuring $\tilde g_{01}$, we used a flux modulation frequency $\omega_p/2\pi = 300~\mathrm{MHz}$ and the frequency excursion needed to bring the qubit 2 in resonance with qubit 1 is $59~\mathrm{MHz}$. In view of this, the renormalization factor is calculated to be  $J_0(\tilde \omega_2/2\omega_p) = J_0(0.0585/(2\times 0.3))= 0.998$.

\begin{table}[]
    \centering
    \begin{tabular}{cccc}
    \hline
    \hline
     Parameters & $q$1 & $q$c & $q$2\\
     \hline
     Max. frequency (GHz) & 3.803 & 5.915 & 3.862\\
     Min. frequency (GHz) & 3.173 & - & 3.207\\
     Anharmonicity (MHz) & 235 & - & 233 \\
     $\sqrt{g_{1c}g_{2c}}$ (MHz) & 92.3 & & 92.3\\
      $g_{12}$ (MHz) & 4.7 &- &4.7\\
     Readout fidelity ($\%$) & 90.1& -& 92.0\\
     $T_{1}(\mu\mathrm{s})$& $70$&-&$56$\\
     $T_{2}^{*}(\mu\mathrm{s})$& $14$&-&$19$\\
     $T_{2}(\mu\mathrm{s})$& $44$&-&$44$\\
     
    \hline
    \hline
    \end{tabular}
    \caption{Measured device parameters. The couplings are obtained by fitting $g_{01}$ versus coupler flux bias data to Eq.~\eqref{gqq}. The coherence times are measured at the dc sweet spot. }
    \label{tab:parameters}
\end{table}

\section{Flux crosstalk}\label{flux_crosstalk}
The qubits and the coupler have a dedicated on-chip, single-ended flux bias line.
In an ideal device each qubit is inductively coupled exclusively to a single flux bias line.
However, because of flux crosstalk at the device level, qubits may be spuriously coupled to multiple flux lines.
The total flux through each SQUID loop is the cumulative effect of a multibias configuration of all flux lines.

In order to exclusively control each qubit using its own dedicated flux bias line,
and generate the relevant offset flux to suppress spurious coupling, we introduce the definition of \textit{operational flux crosstalk}.
The operational flux crosstalk $C_{ij}$ between the $i$th qubit and the $j$th flux bias line is defined as the ratio of the bias currents $I_i$ and $I_j$ that independently generate the same flux through the SQUID loop of the target qubit.
From the definition, it follows that $C_{ij}=\frac{I_i}{I_j}=\frac{m_{ij}}{m_{ii}}$, where $m_{ii}$ and $m_{ij}$ are the direct and spurious mutual inductances between the $i$th qubit and its dedicated flux bias line and the $j$th flux line, respectively.
If currents are measured at the source rather than at the chip level, the inverse of the operational crosstalk matrix $C$ directly provides the multibias currents that effectively suppress the spurious flux on all qubits.

The measurement to extract the operational flux crosstalk involves changing the flux bias on a target qubit $i$ over its period and then simultaneously measuring the frequencies of all resonators.
This measurement is repeated for the flux lines of each of the device resonators.
When qubit $i$ is biased with its own flux line, the resonance frequency versus flux exhibits the usual sinusoidal behavior, from which we extract the reference period $I_{i}$.
When the flux line to another qubit $j$ is biased, the frequency of the resonator $i$ will change depending on the magnitude of the flux crosstalk and therefore exhibits a larger period $I_{j}$.
The magnitude of the operational flux crosstalk between qubit $i$ and flux bias line $j$ is $C_{ij}=I_i/I_j$.
The sign of this crosstalk element is determined from a separate measurement that calculates the relative change of the frequency of the resonator $i$ in response to a change of the current in the flux bias of line $j$.
The measurement is repeated for each qubit and tunable coupler to obtain all elements of the matrix.

Because resonators are measured simultaneously, this method for measuring the operational crosstalk matrix scales with the number of resonators rather than with the number of flux lines. Hence, it is a time-efficient method to measure flux crosstalk even for large-scale quantum processors.

For the device used in this work we measured
\begin{equation*}
C= \left(
    \begin{array}{ccc}
     1 &-0.471 & 0.392\\
     -0.226& 1& 0.248\\
     0.378 &-0.479 &1
\end{array}
\right).
\end{equation*}
written in the basis of $\{$qubit 1, tunable coupler, qubit 2$\}$. We experimentally verified the method by measuring the crosstalk matrix after we applied the correction. We measured an identity crosstalk matrix with error less than $0.01$ for each element of the matrix.

\section{rf flux pulse transfer function}\label{transfer} 
\begin{figure}
    \centering
    \includegraphics[width=\linewidth]{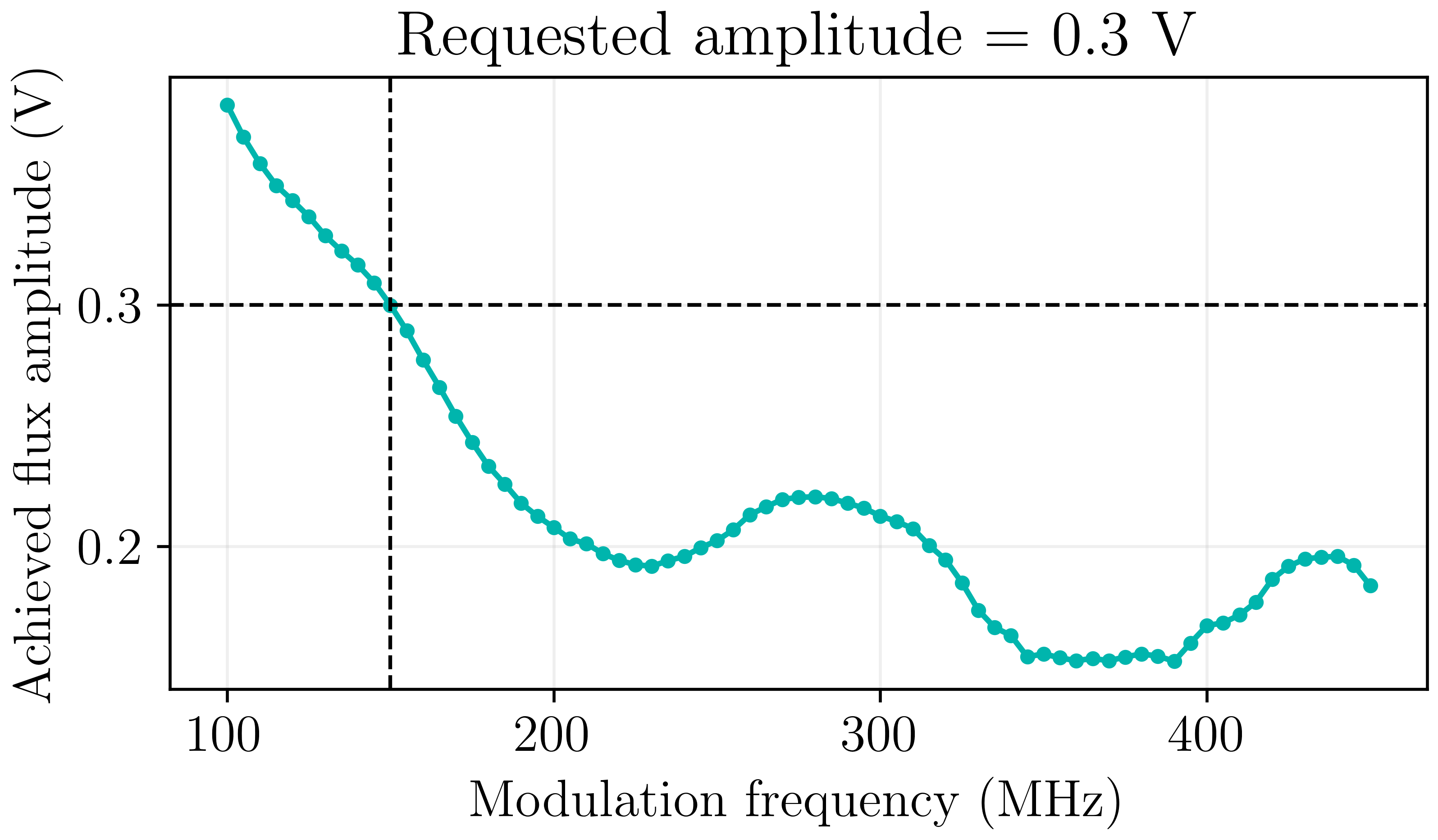}
    \caption{\textbf{Flux pulse transfer function}. Achieved flux modulation amplitude for a given requested amplitude (0.3 V) versus modulation frequency.}
    \label{fig:transfer_function}
\end{figure}

A parametric-resonance gate is implemented by modulating the qubit frequency via a rf flux pulse. Although the gate does not depend on the modulation frequency, the amplitude of the pulse generated by the room-temperature hardware is attenuated substantially at higher modulation frequencies. The dependence of the amplitude of the modulated pulse on the modulation frequency is referred to as the transfer function. We measure the transfer function for the modulated qubit (Fig.~\ref{fig:transfer_function}). The measurement carried out by first calibrating the voltage versus flux qunatum relation by running Ramsey-type experiment where we insert a modulated flux pulse at a given frequency between the $\pi/2$ pulses and measure the excursion of the qubit frequency. We then fix the amplitude of the flux pulse (requested amplitude) and vary the frequency of modulation and measure the excursion of the qubit frequency. At each modulation frequency, we can estimate the amplitude that reached the qubit from the measured frequency excursion and based on the calibration measurement. The transfer function largely depends on the details of the hardware. However, there will be a finite reduction of the amplitude along the fridge lines before it gets to the qubit.  Figure ~\ref{fig:transfer_function} shows a plot of the achieved (at the qubit level) flux modulation amplitude for a given requested amplitude (in this case 0.3 V) versus the modulation frequency. The achieved amplitude significantly drops as the modulation frequency increases. When choosing the modulation frequency to enact parametric resonance gates, it is important to pick a frequency that could give sufficient amplitude to activate various gates.

\section{Repeated iSWAP Quantum Process Tomography fidelity}\label{stability}
\begin{figure}
    \centering
    \includegraphics[width=\linewidth]{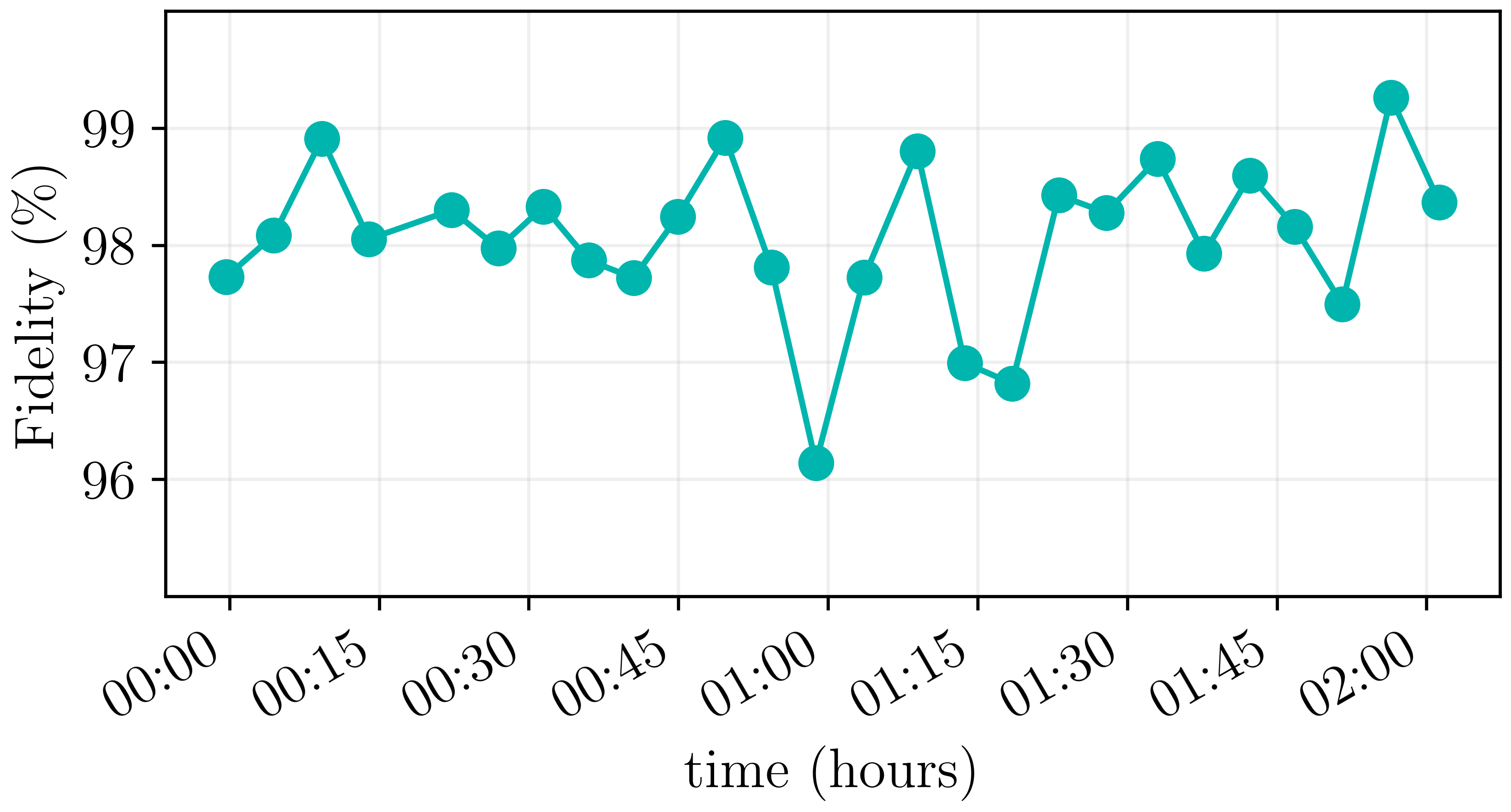}
    \caption{Quantum process tomography fidelity of the iSWAP gate presented in the main text versus time. After each tomography measurement, the virtual Z gates are updated.}
    \label{fig:fidelity_vs_time}
\end{figure}

To study the stability of the gate, we run a repeated quantum process tomography of the iSWAP gate presented in the main text. After each tomography measurement, the virtual Z corrections are updated before the subsequent tomography measurement. The virtual Z correction varies from run to run by a small amount, $<0.05$ rad. The measurement is run over the span of two hours [see Fig.~\ref{fig:fidelity_vs_time}]. The fidelity varies from $96.2\%$ to $99.3\%$, with a median fidelity of $98.2\%$ over the span of two hours. We attribute the fluctuations in the fidelity to the optimal quadrature drift during the measurement because of the hardware's temperature fluctuations. Given that the qubit coherence times $T_1$ are substantially longer than the $T_2^{*}$ during the gate operation, we do not expect the inevitable $T_1$ fluctuations to contribute to fidelity variations. Besides, we have not observed a variation in $T_2^{*}$ over time.

\end{document}